# CubeSat Orbit Insertion Maneuvering Using $J_2$ Perturbation

M. Amin Alandihallaj
University of Luxembourg, 29 Av.
John F. Kennedy, 1855 Kirchberg
Luxembourg
amin.hallaj@uni.lu

M. Reza Emami
Institute for Aerospace Studies,
University of Toronto, 4925 Dufferin
Street, Toronto, Ontario M3H 5T6,
reza.emami@utoronto.ca

*Abstract*— The precise insertion of CubeSats into designated orbits is a complex task, primarily due to the limited propulsion capabilities and constrained fuel reserves onboard, which severely restrict the scope for large orbital corrections. This limitation necessitates the development of more efficient maneuvering techniques to ensure mission success. In this paper, we propose a maneuvering sequence that exploits the natural $J_2$ perturbation caused by the Earth's oblateness. By utilizing the secular effects of this perturbation, it is possible to passively influence key orbital parameters such as the argument of perigee and the right ascension of the ascending node, thereby reducing the need for extensive propulsion-based corrections. The approach is designed to optimize the CubeSat's orbital insertion and minimize the total Δv required for trajectory adjustments, making it particularly suitable for fuel-constrained missions. The proposed methodology is validated through comprehensive numerical simulations that examine different initial orbital conditions and perturbation environments. Case studies are presented to demonstrate the effectiveness of the $J_2$-augmented strategy in achieving accurate orbital insertion, showing a major reduction in fuel consumption compared to traditional methods. The results underscore the potential of this approach to extend the operational life and capabilities of CubeSats, offering a viable solution for future low-Earth orbit missions.

## 1. INTRODUCTION

The utilization of CubeSats for both scientific and technological missions has been growing dramatically since the first CubeSat was launched in 2003 [1]. Their main advantages, including smaller size and mass as well as shorter development time, compared to the traditional satellites, have resulted in an increasing rate of CubeSat launches in recent decades. Since the CubeSats usually utilize Commercial-Off-The-Shelf (COTS) components, employing a CubeSat is also cost-effective for a wide variety of space applications, such as Earth observations[2, 3], communication [4], active orbital debris removal [5], wildfire detection [6], and interplanetary missions [7].

Since CubeSats are launched as secondary payloads, their insertion orbits are guaranteed only within large tolerances. However, it is essential for certain CubeSat missions to put them in precise orbits, especially when having multiple CubeSats in a formation or constellation [8]. Thus, a CubeSat may need to perform some orbital maneuvering, in order to correct its orbit and reach the designated orbital elements.

Due to several key challenges in putting a CubeSat on a precise orbit, orbit insertion maneuvering of CubeSats has not been fully rectified yet. The first challenge is that, since a major part of each CubeSat is often allocated to payloads, a very limited amount of propellant can be carried onboard [9]. Secondly, the propulsion systems currently available for CubeSats offer low-thrust levels, which should be employed for a long duration of time to provide enough *ΔV* needed for orbit corrections [10]. Further, the limited electric power generated onboard CubeSats prevents their propulsion system from being continuously activated for a long time. The third challenge is that an online optimal orbit correction algorithm requires high computational resources, which are typically beyond the processing capabilities of CubeSats [11].

Various techniques have been proposed in the literature to minimize fuel consumption during orbital corrections, including the use of aerodynamic forces [12, 13] and third-body perturbations [14]. While promising, the effectiveness of these methods is highly dependent on specific orbital parameters and the relative positions of celestial bodies such as the Earth, Moon, and Sun. In low-Earth orbit (LEO), where most CubeSats operate [15], a dominant perturbative force is the $J_2$ perturbation, which arises due to the Earth's oblateness. This perturbation, which primarily affects the precession of orbital elements, can potentially be leveraged to reduce the fuel required for orbit correction.

This paper presents a maneuvering algorithm that utilizes the $J_2$ perturbation to enable more efficient CubeSat orbit insertion. By exploiting the secular effects of this perturbation, CubeSats can reduce their reliance on thrust and conserve propellant during orbital corrections. The paper begins by reviewing the governing equations of motion, followed by the formulation of the orbit insertion maneuver using $J_2$ perturbation. Numerical simulations are then conducted to evaluate the effectiveness of the proposed approach, and the results are discussed in detail. The paper concludes by summarizing the findings and outlining future directions for research.

## 2. ORBITAL DYNAMICS

The set of classical orbital elements selected for a satellite state representation includes the semi-major axis, $a$, the eccentricity, $e$, the orbital plane inclination, $i$, the right ascension of ascending node, $Ω$, the argument of perigee, $ω$, and the argument of latitude, $u$. The Gauss' variational equations (GVEs) provide the instantaneous time rates of the orbital elements, considering the secular effect of the $J_2$ perturbation as follows [16-18]:

$$\frac{da}{dt} = \frac{2a^2}{h}\left(e \sin\theta\, f_r + \frac{p}{r} f_c\right) \quad (1a)$$

$$\frac{de}{dt} = \frac{1}{h}\left(p \sin\theta\, f_r + ((p+r)\cos\theta + re)f_c\right) \quad (1b)$$

$$\frac{di}{dt} = \frac{r \cos u}{h} f_n \quad (1c)$$

$$\frac{d\Omega}{dt} = -\frac{3R_e^2 J_2 n}{2p^2}\cos i + \frac{r \sin u}{h \sin i} f_n \quad (1d)$$

$$\frac{d\omega}{dt} = \frac{3R_e^2 J_2 n}{4p^2}(4 - 5\sin^2 i)$$
$$+ \frac{p}{he}\left(\left(1 + \frac{r}{p}\right)\sin\theta\, f_c\right. \quad (1e)$$
$$\left. - \cos\theta\, f_r\right) - \frac{r \sin u}{h \tan i} f_n$$

$$\frac{du}{dt} = \frac{h}{r^2} - \frac{3R_e^2 n J_2}{4p^2}\left(\sin^2 i\left(5 - 3\sqrt{1-e^2}\right)\right.$$
$$\left. + \left(2\sqrt{1-e^2} - 4\right)\right) - \frac{r \sin u}{h \tan i} f_n \quad (1f)$$

In this context, $\theta$ represents the true anomaly, $h$ denotes the orbital angular momentum, $n$ is the mean motion of the orbit, $r$ is the orbital radius, and $p$ refers to the semi-latus rectum of the orbit. Additionally, $J_2$ is the second zonal harmonic coefficient that accounts for the perturbative effects caused by the Earth's oblateness, and $Re$ is the Earth's mean radius. The propulsion acceleration components, $f_r$, $f_c$, and $f_n$, correspond to accelerations in the radial, along-track, and cross-track directions, respectively. These are defined within a local-vertical-local-horizontal (LVLH) reference frame, which is centered on the satellite. This reference frame is commonly used in orbital mechanics to describe the satellite's motion relative to its position and orientation around the Earth.

## 3. ORBITAL MANEUVER THEORY

In-plane maneuvers are required to correct $a$ and $u$, which generally require less $\Delta V$ than out-of-plane maneuvers for correcting $i$ and $\Omega$. The first in-plane maneuver can be defined to modify $a$. Based on the Gauss's Eq. (1a), the $a$ modification maneuver can be derived as:

$$\Delta V_c^a(a_2, a_1) = \frac{hr}{2a_1^2 p}\Delta a \quad (2)$$

where $\Delta V_c = f_c B_t$, and $B_t$ is the propulsion burn time. Note that through the paper the quantities of an orbital element, $x$, before and after the maneuver, $\Delta V^x(x_2, x_1)$, are shown by $x_1$ and $x_2$, respectively, and $\Delta x = x_2 - x_1$.

A double-thrust solution is defined to change $u$ so that $a$ remains the same at the end of maneuver. The pair of along-track maneuvers settles a modified $u$, as follows [19]:

$$\Delta V_c^{u_\pm}(u_2, u_1)$$
$$= \pm\sqrt{\frac{\mu}{a}}\left(\sqrt{2 - \left(1 + \frac{\Delta u}{2\pi k}\right)^{-\frac{2}{3}}} - 1\right) \quad (3)$$

where $k$ denotes the number of revolutions between two thrust impulses. It should be noted that it is assumed that the satellite is in a circular orbit to derive Eq. (3).

According to the Gauss's equations (1c) and (1d), the out-of-plane maneuvers required for modifying $i$ and $\Omega$ can be derived as follows:

$$\Delta V_n^i(i_2, i_1) = \frac{h}{r \cos(u)}\Delta i \quad , u = z\pi \quad (4a)$$

$$\Delta V_n^\Omega(\Omega_2, \Omega_1) = \frac{h \sin i}{r \sin(u)}\Delta\Omega \, , u = \frac{(2z-1)\pi}{2} \quad (4b)$$

where $z$ is an integer number. It should be noted that $\Omega$ and $i$ remain constant during $\Delta V_n^i(i_2, i_1)$ and $\Delta V_n^\Omega(\Omega_2, \Omega_1)$ maneuvers, respectively.

The inclination and the right ascension of ascending node angles can be simultaneously modified using a single out-of-plane maneuver $\Delta V_n^{i\Omega}(\Delta i, \Delta\Omega)$ performed at $u^*$ calculated by equating the right-hand side of Eqs. (4a) and (4b) and solving for $u$, which leads to

$$u^* = \tan^{-1}\left(\frac{\Delta\Omega}{\Delta i}\sin i_1\right) \quad (5)$$

Substituting Eq. (5) into Eqs. (4a) and (4b) and solving for $\Delta V_n^{i\Omega} = \Delta V_n^\Omega = \Delta V_n^i$ gives

$$\Delta V_n^{i\Omega}(\Delta i, \Delta\Omega) = \frac{h}{r}\sqrt{\Delta i^2 + (\Delta\Omega \sin i_1)^2} \quad (6)$$

It should be noted that the general assumption for deriving Eqs. (4)-(6) is that the CubeSat position is fixed during the maneuver. However, if the CubeSat is not equipped with a propulsion system with a sufficiently large thrust to instantaneously change $\Delta i$ and $\Delta\Omega$, the propulsion system needs to operate for a period of time to continuously change the velocity and orbital elements. The variation in $i$ and $\Omega$ using low-thrust maneuvers can be given by

$$\Delta i = f_n \int_0^{B_t} \frac{r \cos u}{h} dt$$
$$\Delta\Omega = f_n \int_0^{B_t} \frac{r \sin u}{h \sin i} dt \quad (7)$$

where $B_t$ is the propulsion burn time. Figure 1 illustrates the relative error of plane change maneuvers using Eqs. (4) with respect to Eq. (7) as a function of $B_t$ for a sun-synchronous 786 km orbit, where it is shown that although the relative error increases gradually as $B_t$ increases, the relative error is still less than 1% for $B_t \leq 470s$. It should be noted that if a small maximum burn time $(B_{t_{max}})$ is selected to reduce the error, multiple consecutive maneuvers will be required to achieve the desired changes in the velocity and orbital elements. For this purpose, $u^*$, $\Delta V_n^{i\Omega}(\Delta i, \Delta\Omega)$, and the required burn time to provide $\Delta V_n^{i\Omega}(\Delta i, \Delta\Omega)$, $B_{t_{required}} = \Delta V_n^{i\Omega}(\Delta i, \Delta\Omega)/f_n$, should be calculated using Eqs. (5-6). Then the propulsion system applies a thrust with a burn time equal to $\min\left(B_{t_{max}}, B_{t_{required}}\right)$. If $B_{t_{required}} > B_{t_{max}}$, the new $u^*$ and $\Delta V_n^{i\Omega}(\Delta i, \Delta\Omega)$ are calculated in the next revolution for the next thrust.

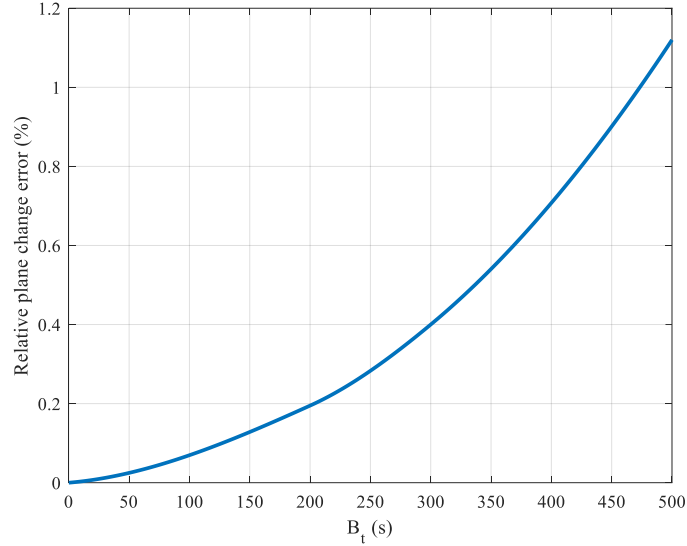

**Figure 1 The relative plane change error with respect to the burn time.**

## 4. ORBITAL MANEUVER PLANNING

In this section, two maneuvering sequences are proposed for transferring a CubeSat into its designated final orbit. The first sequence follows a conventional approach, applying the maneuvers discussed in the preceding section for orbit correction, as outlined by Eqs. (2-6). This method does not take advantage of the $J_2$ perturbation effects. For clarity throughout the remainder of the paper, the orbital elements corresponding to the initial, transfer, and final orbits will be denoted with the subscripts "0", "t", and "f", respectively. Additionally, it is assumed that each maneuvering sequence begins at the initial time epoch $t_0$ and concludes at $t_0 + \tau$, where $\tau$ represents the total maneuvering duration.

The classic maneuvering sequence begins by simultaneously modifying the right ascension of the ascending node ($\Omega$) and the inclination ($i$) through a combined normal maneuver, denoted as $\Delta V_n^{i\Omega}(\Delta i, \Delta\Omega)$, where the changes in inclination and right ascension are $\Delta i = i_f - i_0$ and $\Delta\Omega = \Omega_f - \Omega_0$, respectively. Following this, an in-plane maneuver, $\Delta V_c^a(a_f, a_0)$, is performed to adjust the semi-major axis ($a$), where the target value is $a_f$ and the initial value is $a_0$.

Finally, the sequence concludes with a maneuver $\Delta V_c^{u\pm}(u_f, u_1)$, which is designed to adjust the argument of latitude ($u$) to its desired value $u_f$, where $u_1$ represents the argument of latitude just before the initiation of this maneuver. It is important to note that due to the change in semi-major axis between the two impulses of the $\Delta V_c^{u\pm}(u_f, u_1)$ maneuver, the rate of change of the right ascension of the ascending node $\dot\Omega$ will be affected by the $J_2$ perturbation. Consequently, the final value of $\Omega$ may differ from the desired value $\Omega_f$, requiring an additional maneuver to correct the right ascension, denoted as $\Delta V_n^{\Omega}(\Omega_2, \Omega_1)$.

The drift in $\Omega$, denoted as $\Delta\Omega_e$, that occurs during the $\Delta V_c^{u\pm}(u_f, u_1)$ maneuver can be approximated using a first-order Taylor series expansion of Gauss's equation (Eq. 1d) as follows

$$\Delta\Omega_e = \frac{21 R_e^2 J_2 n_f}{4 a_f^3} \cos(i_f) \Delta t_t \qquad (8)$$

where $\Delta t_t$ denotes the time duration between the pair of $\Delta V_c^{u\pm}(u_f, u_1)$ impulses. This approximation helps quantify the deviation in $\Omega$ due to the perturbation, which must be compensated in the final orbital adjustments. Thus, the total required $\Delta V$ for the classic sequence can be obtained as

$$\Delta V_C = \left|\Delta V_n^{i\Omega}(\Delta i, \Delta\Omega)\right| + \left|\Delta V_c^a(a_f, a_0)\right| \\ + \left|\Delta V_c^{u\pm}(u_f, u_1)\right| + \left|\Delta V_n^{\Omega}(\Delta\Omega_e)\right| \qquad (9)$$

In contrast to the classic approach, the $J_2$-enhanced maneuvering sequence seeks to exploit the $J_2$ perturbation to assist in orbital corrections. This method involves placing the CubeSat into a transfer orbit characterized by an appropriate semi-major axis $a_t$ and inclination $i_t$, chosen such that the $J_2$ perturbation gradually reduces the errors in the right ascension of the ascending node ($\Omega$) and the argument of latitude ($u$) relative to the desired orbit over the maneuvering duration.

If the initial orbit and the transfer orbit share the same right ascension of the ascending node at the initial epoch, such that $\Omega_t = \Omega_0$, the necessary condition for the convergence of $\Omega$ to its target value, while utilizing the $J_2$ perturbation, can be expressed as follows:

$$\dot{\Omega}_f = \dot{\Omega}_0 + \frac{\Omega_f - \Omega_0}{\tau} \quad (10)$$

which can be substituted into the Gauss's Eq. (1d) to find $i_t$ as

$$i_t = \cos^{-1}\left(\frac{-\dot{\Omega}_f + \frac{\Delta\Omega}{\tau}}{3R_e^2 J_2 n_t} 2p_t^2\right) \quad (11)$$

The necessary condition to achieve the desired value of argument of latitude, $u_f$ is

$$a_t = a_f \sqrt[3]{\left(1 + \frac{(u_f - u_0) + \delta u_{j2}}{2\pi k}\right)^2} \quad (12)$$

Here, $k$ denotes the number of revolutions the CubeSat completes in the transfer orbit. The variation in the argument of latitude, $\delta u_{j2}$, due to the $J_2$ perturbation, can be approximated using a first-order Taylor series expansion of Eq. (1f), as follows:

$$\delta u_{j2} = -\frac{3R_e^2 J_2}{a_f^3}(\sin^2 i_f - 1)\left(\frac{3\mu}{4a_f^3 n_f} + n_f\right)(a_f - a_t)\tau - \frac{3R_e^2 n J_2}{2a_f^2}\sin(2i_f)(i_f - i_t)\tau \quad (13)$$

Figure 2 illustrates the sequence of maneuvers for the $J_2$-enhanced orbit correction strategy. The sequence starts with an in-plane maneuver $\Delta V_c^a(a_t, a_0)$, which adjusts the semi-major axis from the initial value $a_0$ to the transfer orbit value $a_t$. Following this, a pure inclination change maneuver $\Delta V_n^i(i_t, i_0)$ is executed to achieve the inclination of the transfer orbit $i_t$.

Once the CubeSat is in the transfer orbit, the $J_2$ perturbation gradually causes the argument of latitude (u) and the right ascension of the ascending node ($\Omega$) to drift towards their desired values. During this phase, the CubeSat passively adjusts its trajectory under the influence of the Earth's oblateness. After the argument of latitude aligns with the target value, an additional in-plane maneuver $\Delta V_c^a(a_f, a_t)$ is performed during the final orbital revolution of the maneuvering period, bringing the semi-major axis to its final value $a_f$.

Finally, a pure inclination change maneuver $\Delta V_n^i(i_f, i_t)$ is conducted at time $t = t_0 + \tau$, completing the orbital insertion by achieving the desired inclination $i_f$. It is important to note that because the out-of-plane maneuvers are performed when the argument of latitude $u = z\pi$, these maneuvers do not affect the argument of latitude. The graphical schematic diagram is illustrated in Figure 3.

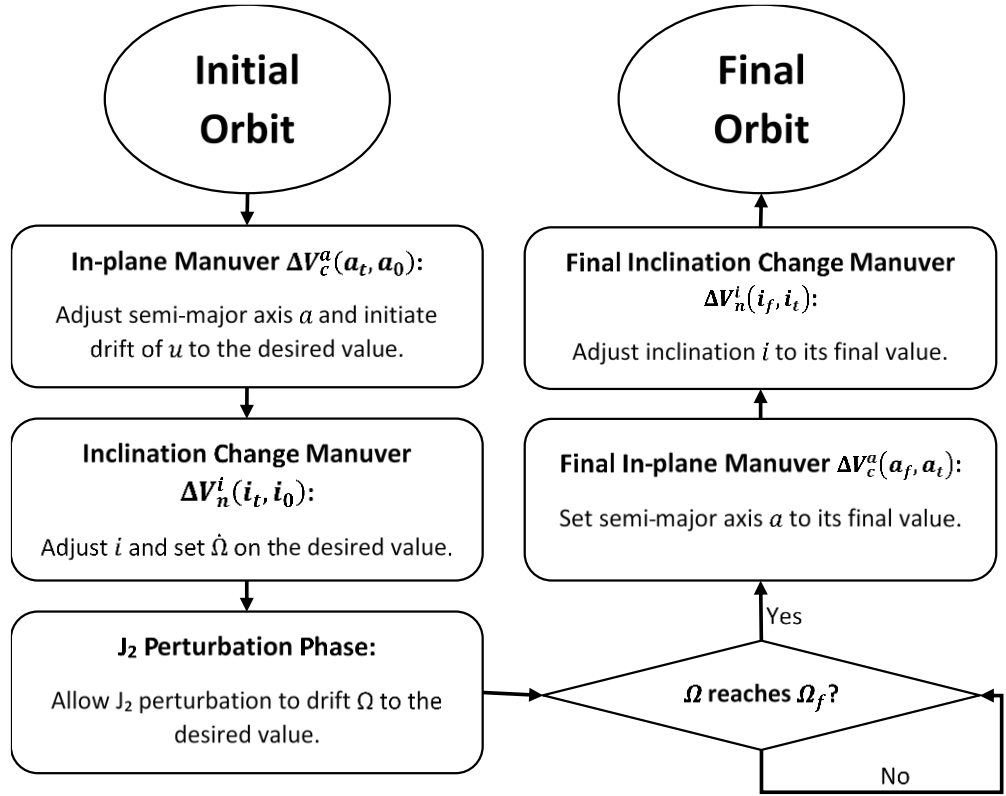

**Figure 2 The Block diagram of the $J_2$ maneuvering sequence.**

The total ΔV required for this sequence can be calculated as:

$$\Delta V_J = |\Delta V_c^a(a_t, a_0)| + |\Delta V_n^i(i_t, i_0)| + |\Delta V_c^a(a_f, a_t)| + |\Delta V_n^i(i_f, i_t)| \quad (14)$$

## 5. NUMERICAL SIMULATION

The performance of the proposed orbital maneuvering sequence is investigated through an orbit correction scenario for a 10kg CubeSat equipped with a 0.1N continuous thruster with the maximum burn time of 450 seconds for each maneuver. It is also assumed that $t_0 = 2$ days and $\tau = 30$ days. The initial deviations of orbital elements with respect to the desired circular sun-synchronous 786 km orbit for two cases are listed in Table 1.

**Table 1 The orbital elements deviation between the initial and desired orbits**

| Case | $\delta\Omega$ (Deg.) | $\delta i$ (Deg.) | $\delta a$ (km) | $\delta u$ (Deg.) |
|------|------|------|------|------|
| A | -0.5 | 0.1 | -10 | 180 |
| B | 0.1 | 0.05 | 1 | 10 |

Figure 4 compares the evolution of deviations in the orbital elements and the distance $L$ between the real and desired CubeSat position, using the classic and $J_2$ sequences for Case A. It can be seen from the figure that with the $J_2$ sequence, $\delta i$ increases at the epoch to provide sufficient rate of $\delta\Omega$, so that $\delta\Omega$ is converged to zero at a constant rate by the end of the sequence, while the CubeSat maintains its inclination $i_t$. It can also be observed that $\delta a$, $\delta e$, and $\delta u$ are concurrently reduced to zero at the end of the sequence. It is also shown that the CubeSat gradually approaches to the desired position, and the desired position is maintained with an accuracy of better than 1.2 km. The final error in the CubeSat position is caused by Taylor series truncation in the Gauss's equations as well as the fact that the implementation of thrust is continuous not impulsive. As for the classic sequence, the figure shows that $i$ and $\Omega$ are modified concurrently at the epoch. However, because of the $J_2$ perturbation, $\delta\Omega$ increases slowly, which is compensated by using $\Delta V_n^\Omega(\Delta\Omega_e)$ at the end of the classic sequence. Moreover, it is shown that the position is maintained with an error of better than 2 km with respect to the desired position. The reason why final $\delta L$ in the classic sequence is larger than that of $J_2$ sequence is that the last maneuver of the classic sequence ($\Delta V_n^\Omega(\Delta\Omega_e)$) for correcting $\Omega$ also changes $u$ (slightly), which makes the CubeSat drift from the desired position. The last maneuver for correcting $\Omega$ is needed, because during the maneuver for correcting $u$ both $i$ and $\Omega$ are presumed to remain unchanged, but $\Omega$ practically changes due to the $J_2$ perturbation. Consequently, the longer the $u$ correction maneuver takes, the more correction is needed for $\Omega$ at the end, resulting in more deviation in $u$ eventually, which in turn causes more error in the final position. In the $J_2$ sequence, on the other hand, since the correction in $\Omega$ occurs by the perturbation itself, given sufficient time, $u$ can eventually approach its desired value, resulting in a more accurate position.

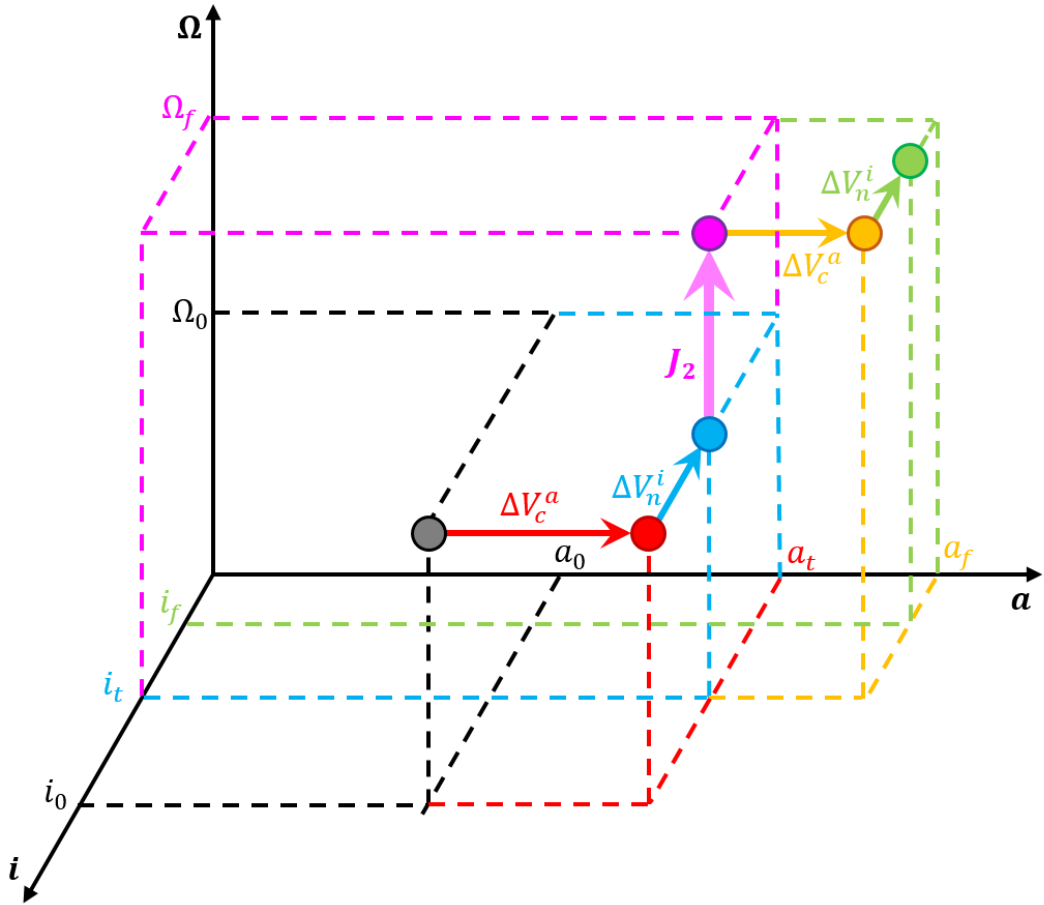

**Figure 3 Schematic diagram of the $J_2$ maneuvering sequence.**

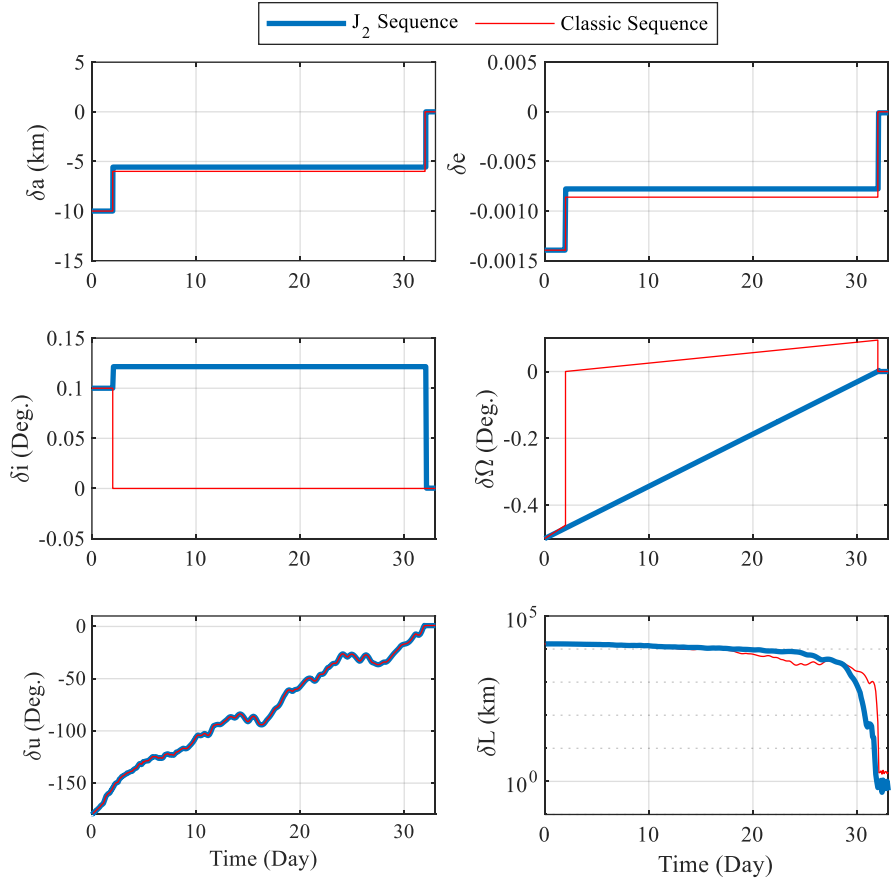

**Figure 4 The time histories of evolution of errors in the orbital elements using the classic and $J_2$ sequences.**

In addition to the accuracy of final position, the $J_2$ sequence outperforms the classic sequence in terms of fuel consumption for case A. The required $\Delta V$ for the $J_2$ sequence is 24 m/s, whereas it is 91 m/s for the classic sequence. However, one cannot conclude that the $J_2$ sequence always provides a solution with a lower fuel consumption. Figure 5 and Figure 6 depict the required $\Delta V$ for Cases A and B, respectively, using the classic and $J_2$ sequences as a function of the period of sequence $\tau$. It can be observed that the classic sequence provides a solution with less $\Delta V$ than $J_2$ sequence in Case A when $\tau$ is less than 9 days, and in Case B when $\tau$ is less than 24 days. It can also be observed that although the required $\Delta V$ generally decreases as $\tau$ increases, $\Delta V$ for Case A using $J_2$ sequence remains nearly constant for $\tau > 37$ days. The reason is that $i_0 \leq i_t \leq i_f$ for $\tau > 37$ days. Thus $|\Delta V_n^i(i_t, i_0)| + |\Delta V_n^i(i_f, i_t)| = |\Delta V_n^i(i_f, i_0)|$ remains constant. In this case $i_t = i_0$ when $\tau = 37$ days, thus the rate of $\delta\Omega$ is sufficient for converging to zero at the end of sequence. If $\tau < 37$ days, as $\tau$ decreases, $|i_f - i_t|$ will become larger, which consequently increases $|\Delta V_n^i(i_f, i_t)|$. In case B, where $ii_f < i_t$, $\Delta V$ decreases with the J$_2$ sequence, reaching a minimum as $\tau$ approaches infinity. If the maneuvering period is limited, the J$_2$ sequence increases fuel consumption, making the classic sequence preferable. However, with a longer maneuvering period, $J_2$ perturbation can help reduce fuel use.

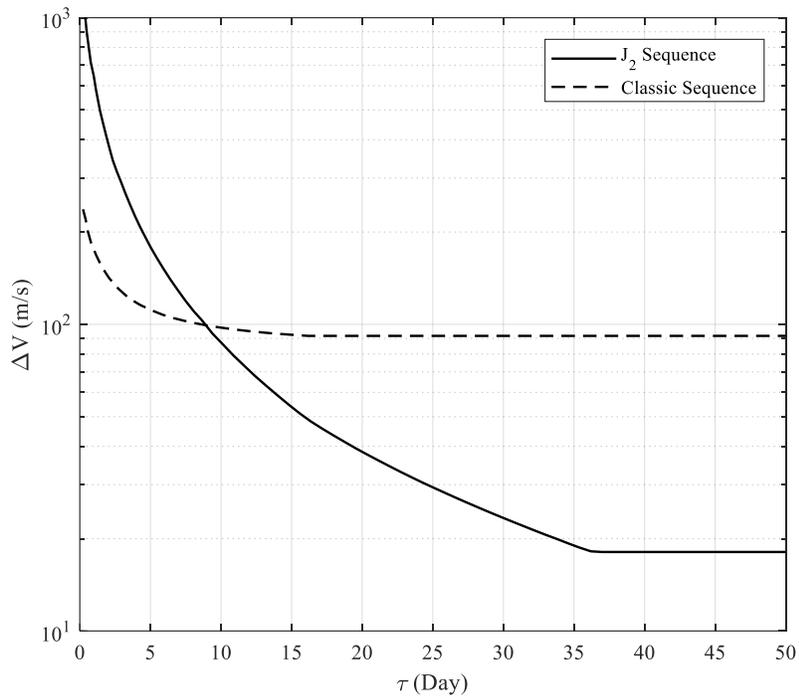

**Figure 5** The comparison of the required $\Delta V$ using the classic and $J_2$ sequences for Case A.

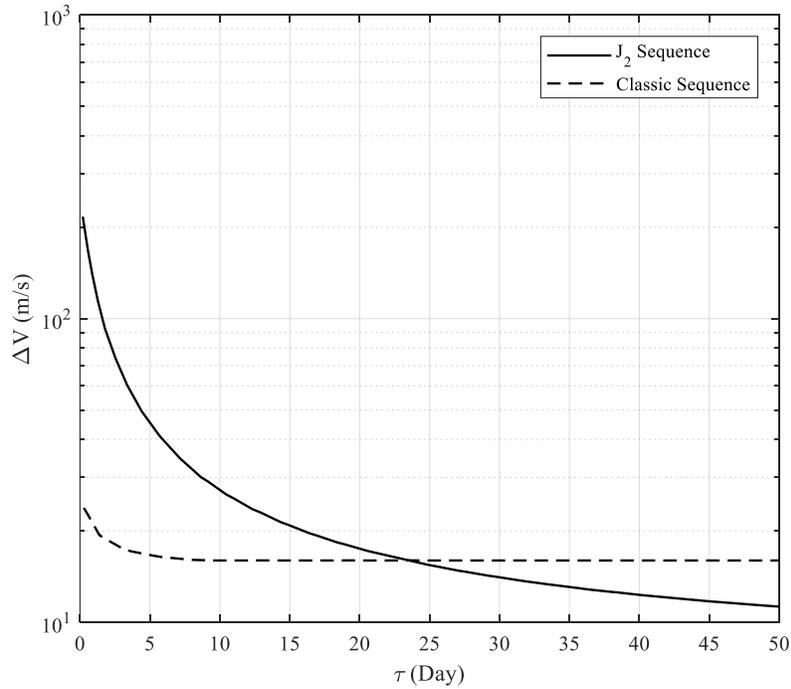

**Figure 6** The comparison of the required $\Delta V$ using the classic and $J_2$ sequences for Case B.

As demonstrated in the case studies, the efficiency of each maneuvering sequence depends on the specific conditions, which are primarily a combination of the initial errors in inclination ($i$) and right ascension of the ascending node ($\Omega$), as well as the available maneuvering time ($\tau$). Each sequence can outperform the other depending on these factors.

Figure 7 illustrates the relative efficiency of the $J_2$ Sequence compared to the classic sequence. The relative efficiency is quantified using the parameter $\rho = 1 - \left(\frac{\Delta V_J}{\Delta V_C}\right)$. This figure plots $\rho$ as a function of the maneuvering period ($\tau$) for varying initial errors in two other orbital angles.

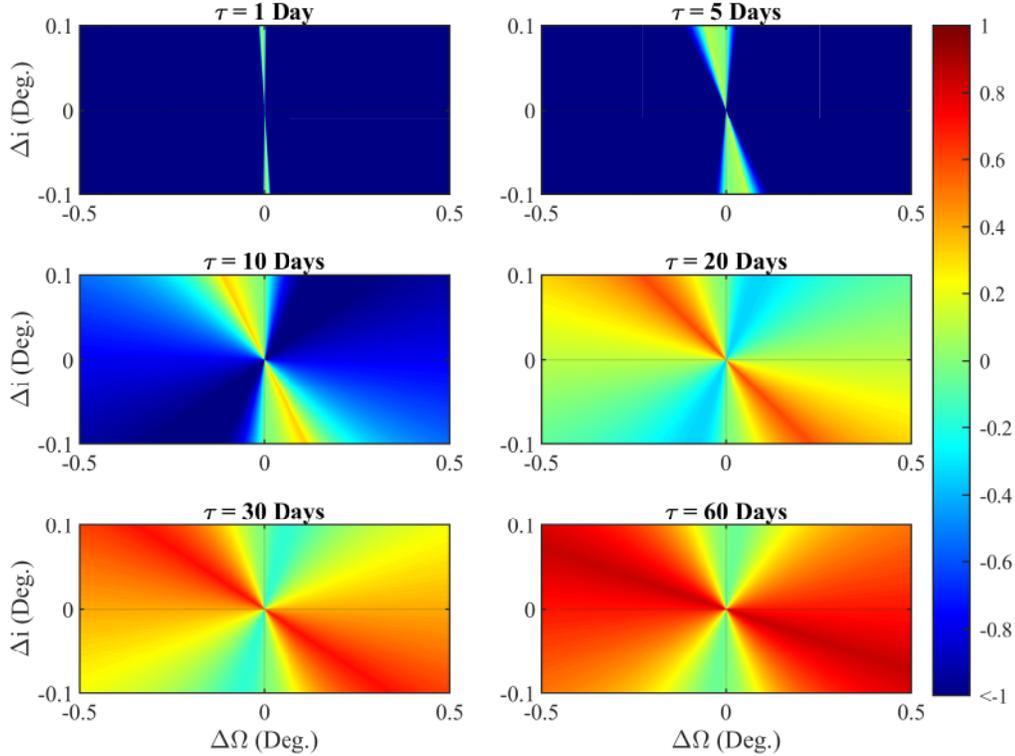

**Figure 7** The efficiency of the $J_2$ sequence with respect to the classic sequence.

For this particular case study, the remaining orbital parameters—such as the semi-major axis and eccentricity—are assumed to be accurate at the time of orbital insertion. Therefore, their influence on the required $\Delta V$ is not considered in this analysis. In the figure, regions depicted in shades of blue indicate areas where the classic maneuvering sequence requires less $\Delta V$ than the $J_2$ sequence, highlighting that the classic sequence is more efficient under those conditions.

It can be observed from the figure that as the maneuvering period ($\tau$) increases, the $J_2$ sequence becomes increasingly efficient, with fuel savings reaching up to 85%. This is because the $J_2$ perturbation has more time to exert its influence on the orbital elements, allowing for a more passive correction of the trajectory and reducing the need for active propulsion. However, for shorter maneuvering periods, the classic sequence proves to be more effective, as it can achieve the necessary orbital corrections more quickly with fewer constraints on the time available for the perturbation to take effect.

The $J_2$ sequence is beneficial for missions where longer maneuvering durations are permissible, enabling significant fuel savings by leveraging natural perturbative forces. On the other hand, the classic sequence is better suited for scenarios requiring rapid adjustments, where time is limited, and quick maneuvering is more critical than maximizing fuel efficiency.

## 6. SUMMARY

An orbital maneuvering sequence was developed in the paper for inserting a newly launched CubeSat into the designated orbit, considering the limitations on available propulsion systems and power and processing capabilities for CubeSats. The sequence benefits from the $J_2$ perturbation to reduce fuel consumption by putting the CubeSat on a proper trajectory with a sufficient rate of right ascension of ascending node due to the $J_2$ perturbation. Utilizing the beneficial effects of the $J_2$ perturbation eliminates the need for performing essential plane change maneuvers to modify the right ascension of ascending node, to which a major part of the fuel consumption is dedicated. The effectiveness of the proposed approach is investigated through numerical simulations. The obtained results demonstrated that not only is the proposed sequence superior to the classic sequence in terms of accuracy, but also it is also possible to save considerable $\Delta V$ by employing the $J_2$ perturbation effect for the maneuvering sequence.

## BIOGRAPHY

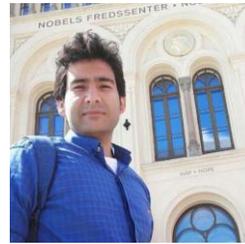

***Dr. M. Amin Alandihallaj*** is the Head of the CubeSat Laboratory and a Research Associate at the Interdisciplinary Centre for Security, Reliability, and Trust (SnT) at the University of Luxembourg. He earned his Ph.D. in Aerospace Engineering with a specialization in controlling space systems from Sharif University of Technology in 2018. Prior to that, he completed both his M.Sc. and B.Sc. in Aerospace Engineering from the same institution. Dr. Alandihallaj furthered his academic career as a Postdoctoral Researcher at the University of Toronto, where his research focused on designing distributed space systems and satellite formations. In 2022, he joined the University of Luxembourg's SnT, where he leads pioneering research on CubeSat technologies, specifically in the areas of space system design and control for small satellites.

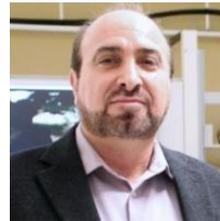

***Prof. M. Reza Emami*** is the Director of the Aerospace Mechatronics Group and Coordinator of the Aerospace and Design Laboratories at the University of Toronto Institute for Aerospace Studies. He was also the Founding Chaired Professor of Onboard Space Systems and Director of Asteroid Engineering Centre at the Luleå University of Technology, Sweden, in 2015-2020. He obtained his Ph.D. and postdoctoral fellowship in Robotics and Mechatronics from University of Toronto, and M.Sc. and B.Sc. in Aerospace Engineering from Sharif University of Technology, and worked in the industry as a licensed Professional Engineer and Project Manager before joining the academia. His research focuses on concurrent engineering of multidisciplinary systems, such as spacecraft, robot manipulators and rovers. Some of his current research includes: space systems miniaturization, space debris mitigation and remediation, concurrent base-arm control of free-flying space manipulators, satellite formation flying, asteroid redirection for exploration and mining, intelligent heterogeneous rover/satellite teams, and reconfigurable manipulators. He has been nominated for several awards, including National Innovation for Technology Award and OPAS Award for Excellence in Teaching. He was also recognized by MathWorks as a Featured Professor in 2013 and 2003, and was in the list of University of Toronto's Miracle Workers in 2000.